\documentstyle[aps,prl,epsf,twocolumn]{revtex}
\begin{document}
\draft
\title{Intermediate mass strangelets are positively charged}
\author{Jes Madsen}
\address{Institute of Physics and Astronomy, University of Aarhus, 
DK-8000 \AA rhus C, Denmark}
\date{August 18, 2000; To appear in Phys.Rev.Lett.\ {\bf 85} p.~4687 (27 Nov 2000)}
\maketitle

\begin{abstract}
For a limited range of parameters, stable strange quark matter may 
be negatively charged in bulk due to one gluon exchange interactions. 
However, the reduction in strange quark occupation
in the surface layer, which is responsible for surface tension,
more than compensates this for intermediate mass 
strangelets, which therefore always have
positive quark charge (e.g.\ for 
baryon number between $10^2$ and $10^{18}$ assuming $\alpha_S=0.9$). 
While details are sensitive to the choice of renormalization, 
the general conclusion is not. This rules out a scenario
where negatively charged strangelets produced in ultrarelativistic 
heavy ion colliders might
grow indefinitely with potentially disastrous consequences.
\end{abstract}


There has recently been some concern about the possibility that a 
negatively charged strangelet formed in ultrarelativistic heavy ion 
collisions might grow by absorbing nuclei to, in principle, 
swallow the Earth. Two investigations have described a number of
reasons, theoretical as well as experimental, why such a scenario is 
exceedingly unlikely\cite{jaffe00}.

Many unlikely coincidences are required for a dangerous situation to
develop. Positively charged quark matter repels ordinary nuclei, so at
the very least, the small strangelets 
(baryon number $A\ll 4\times 10^2$, where $4\times 10^2$ is the total number 
of nucleons involved in a collision of two gold
or lead nuclei) hypothetically formed in
ultrarelativistic heavy ion collisions must have negative quark charge.
And for growth to proceed, not only must the ultimate state of stable bulk
strange quark matter be negatively charged, but so must the path of
intermediate mass strangelets all the way from low $A$ to
$A\rightarrow\infty$.

The present study demonstrates, that even in the unlikely case 
where bulk strange quark matter is both stable and has negative quark charge, 
and where a long-lived, negatively charged strangelet is formed in a collision, 
such a strangelet {\sl cannot} grow significantly by
absorbing nuclei. The reason is, that the quark charge of
intermediate mass strangelets near
the lowest energy state is always positive regardless of the bulk charge.
Thus the potentially dangerous strangelet growth path is blocked by
Coulomb repulsion.

Following the original suggestions of Bodmer\cite{bodmer71} 
and Witten\cite{witten84},
Farhi and Jaffe \cite{farhi84} presented a detailed study of strange quark 
matter properties in bulk, as well as finite size strangelets. 
Concerning the issue of the electric charge of bulk strange matter they 
showed, that the quark charge of stable strange matter is always positive 
for small strong interaction constant, $\alpha_S$ \cite{charge}.
This is easy to understand physically, because of the fortuitous 
cancellation of charge for a gas with equal numbers of up, down and 
strange quarks. For finite strange quark mass and up
and down quark masses close to zero, the number of $s$-quarks will be 
reduced, and the net quark charge be positive, even though the actual 
numbers are small, with a typical charge-to-baryon number ratio of 
$10^{-3}$--$10^{-7}$. One gluon exchange interactions are repulsive for 
massless up and down quarks, but attractive for massive,
nonrelativistic quarks. Thus, for sufficiently high $\alpha_S$ 
Farhi and Jaffe showed, that there is a part of parameter space where 
this effect increases the abundance of $s$-quarks
enough to give the system a total negative quark charge. 
Farhi and Jaffe noted, that details were quite sensitive to the choice 
of renormalization point, and speculated that it might not be
stable with respect to including higher orders in $\alpha_S$. 
But negative quark charge for
bulk strange quark matter seems a valid albeit unlikely possibility.

The energy and quark content of finite size lumps of strange quark matter 
(sometimes called strangelets, at least in the low-$A$ regime) is 
modified relative to that of bulk strange quark matter. 
These finite size effects (surface tension and curvature energy) generally
destabilize strangelets
\cite{farhi84,berger87,madsen93a,gilson93,madsen93b,madsen94}, 
but they also change the overall charge of the system, and as shown 
below they always lead to a wide range of intermediate mass strangelets 
with positive rather than negative charge. Physically 
the main effect of the finite size is a depletion of $s$-quarks in 
the surface layer, which removes negative charge from the system 
relative to the bulk solution. The underlying mechanism is a
simple consequence of quantum mechanics. Whereas a relativistic,
low-mass quark may have nonzero density at the system boundary, the wave
function (and therefore the density) must be zero in the nonrelativistic
limit of a very massive quark. The massive $s$-quarks are therefore 
suppressed more than $u$ and $d$ near the surface\cite{nuclear}.

It will be shown below, that finite size effects increase the total
charge per baryon by an amount
$\Delta Z/A \approx 0.3 A^{-1/3}$. This number is to be
compared with a negative charge per baryon (numerically) less than $10^{-6}$ 
for stable bulk quark matter with $\alpha_S=0.9$, and $10^{-3}$ for
$\alpha_S=1.2$. Thus the total charge of strangelets becomes positive
for $A<10^{18}$ and $A<10^7$ respectively.

In the ideal Fermi-gas approximation the energy of a system composed of
quark flavors $i$ is given by
\begin{equation}
E=\sum_i(\Omega_i+N_i\mu_i)+BV .
\label{Estrangelet2}
\end{equation}
Here $\Omega_i$, $N_i$ and $\mu_i$ denote thermodynamic potentials,
total number of quarks, and chemical potentials, respectively. $B$ is
the bag constant, $V$ is the bag volume. (For simplicity, all
thermodynamical equations are
written without explicit electron/positron terms. Such terms
were included in the numerical calculations).

In the multiple reflection expansion framework of Balian and Bloch
\cite{balian70}, the
thermodynamical quantities can be derived from a density of states of
the form
$
{{dN_i}\over{dk}}=6 \left\{ {{k^2V}\over{2\pi^2}}+f_S\left({m_i\over
k}\right)kS+f_C\left({m_i\over k}\right)C+ .... \right\} ,
$
where area $S=4\pi R^2$ and curvature $C=8\pi R$ for a sphere.
The volume term is universal, whereas the 
functions $f_S$ and $f_C$ depend on the boundary conditions.
The number of quarks of flavor $i$ is
$
N_i=\int_0^{k_{Fi}}({dN_i}/{dk})dk=n_{i,V}V+n_{i,S}S+n_{i,C}C,
$
with Fermi momentum $k_{Fi}=\mu_i(1-\lambda_i^2)^{1/2}$;
$\lambda_i\equiv m_i/\mu_i$.
The corresponding thermodynamical potentials are
$\Omega_i=\Omega_{i,V}V+\Omega_{i,S}S+\Omega_{i,C}C$ ,
where $\partial\Omega_i/\partial\mu_i=-N_i$, and
$\partial\Omega_{i,j}/\partial\mu_i=-n_{i,j}$. 
Minimizing the total energy at fixed $N_i$
gives the energy for a spherical quark lump as
$E=\sum_i N_i\mu_i+(1/3)\Omega_{i,S}S+(2/3)\Omega_{i,C}C.$

In the following I will assume MIT bag model boundary conditions
\cite{degrand75}, but
the main conclusion depends only on the fact that $f_S$ is negative.

The volume terms including first order $\alpha_S$ corrections are given by
\begin{eqnarray}
\Omega&&_{i,V}=-{{\mu_i^4}\over {4\pi^2}}\left( (1-\lambda_i^2)^{1/2}(1-
{5\over 2}\lambda_i^2)
+{3\over 2}\lambda_i^4\ln{{1+(1-\lambda_i^2)^{1/2}}\over\lambda_i}\right.\cr
&&\left. -{{2\alpha_S}\over{\pi}}\left[ 3\left[ (1-\lambda_i^2)^{1/2}-
\lambda_i^2\ln (1+(1-\lambda_i^2)^{1/2})\right]^2 \right.\right.\cr
&&\left.\left. -2(1-\lambda_i^2)^2
-3\lambda_i^4\ln^2\lambda_i \right.\right.\cr
&&\left.\left. +6\ln\left( {{\sigma} \over
{\mu_i}}\right) \left( \lambda_i^2(1-\lambda_i^2)^{1/2}- \lambda_i^4
\ln \left({{1+(1-\lambda_i^2)^{1/2}}\over{\lambda_i}}\right) \right)
\right] \right) ,
\end{eqnarray}
and $n_{i,V}=-\partial\Omega_{i,V}/\partial\mu_i$. 
The expression for $\Omega_{i,V}$ is taken from \cite{farhi84} apart
from correction of a sign misprint. Farhi and Jaffe discuss at length
the choice of renormalization scale, $\sigma$. In accordance with their
work I have picked a value of $\sigma=313$MeV in the following. Details
of the numerical results are sensitive to this choice, especially at
high $\alpha_S$, but calculations for other choices of $\sigma$
assure that the conclusions are invariant.

The surface contribution follows from
$f_S(m/k)=-\left[ 1-(2/\pi)\tan^{-1}(k/m)\right] /8\pi$\cite{berger87}
\begin{eqnarray}
\Omega_{i,S}&&={3\over{4\pi}}\mu_i^3\left[{{(1-\lambda_i^2)}\over 6}
-{{\lambda_i^2(1-\lambda_i )}\over 3}\right.\cr
&&\left. -{1\over{3\pi}}\left(\tan^{-1}\left[
{{(1-\lambda_i^2)^{1/2}}\over\lambda_i}\right]-2\lambda_i (1-\lambda_i^2)^{1/2}
\right.\right.\cr
&&\left.\left.
+\lambda_i^3\ln\left[{{1+(1-\lambda_i^2)^{1/2}}\over\lambda_i}\right]\right)
\right] ;
\end{eqnarray}
\begin{eqnarray}
n_{i,S}&&=-{3\over{4\pi}}\mu_i^2\left[{{(1-\lambda_i^2)}\over 2}
-{1\over{\pi}}\left(\tan^{-1}\left[
{{(1-\lambda_i^2)^{1/2}}\over\lambda_i}\right]
\right.\right.\cr
&&\left.\left.
-\lambda_i (1-\lambda_i^2)^{1/2}
\right)\right] .
\end{eqnarray}
For massless quarks $\Omega_{i,S}=n_{i,S}=0$. For massive quarks
$\Omega_{i,S}$ (which is always positive, except for $\lambda_i=0$ or 1)
has the properties of a surface tension. The corresponding change in
quark number per unit area, $n_{i,S}$, is always negative, approaching 0
for $\lambda_i\rightarrow 0$ (massless quarks) or 1 (infinitely massive
quarks, i.e.\ no $i$-quarks present). 

Curvature terms ($\Omega_{i,C}$ and $n_{i,C}$) are decisive for
understanding the average physical properties of small $A<100$
strangelets, but are not very important for the results at higher $A$.
They were selfconsistently included in
the numerical calculations, using the expressions from
\cite{madsen94}. 
Coulomb energies are negligible because of the
low charge-to-mass ratio.

Figures 1--4 show results of applying the above equations to strange
quark matter at $\alpha_S=0.9$, varying $B$ and $m_s$, 
and assuming massless $u$- and $d$-quarks.
Figure 1 shows contour curves for equal energy per baryon of bulk
strange quark matter in chemical equilibrium, i.e.\ with $\mu_d=\mu_s$
maintained mainly by $u+d\leftrightarrow s+u$, and $\mu_d=\mu_u+\mu_e$,
normally maintained by reactions like $d\rightarrow u+e^-
+\bar\nu_e$. Typical quark chemical potentials are 300 MeV,
whereas $\mu_e$ is a few MeV. A special situation occurs when
$-m_e<\mu_e<m_e$, since here the weak reactions involving electrons or
positrons are blocked, and the total quark charge of the system is zero
with neither electrons nor positrons present. For $\mu_e>m_e$ local
charge neutrality is maintained by a small fraction of electrons,
whereas positrons take over for $\mu_e<-m_e$. Dotted curves in 
Fig.~1 are contours for fixed number of electrons per baryon (positrons
counted as negative number of electrons), which equals the net quark
charge per baryon, $Z/A$. A characteristic property of strange quark
matter is a very low $Z/A$, except for very high $m_s$, where $s$-quarks
are energetically unfavorable. For $-m_e<\mu_e<m_e$, $Z/A$ becomes 0, but for
low $\alpha_S$ this happens only for unrealistically low $m_s$, and
negative bulk charge does not occur at all. For high $\alpha_S$, the
regime of zero charge propagates to intermediate $m_s$ values, and a
region of negative bulk $Z/A$ shows up. For $\alpha_S=0.9$ the bulk
$Z/A$ is still numerically very small, reaching only values of $-10^{-6}$,
whereas $Z/A\approx -10^{-3}$ may be reached for a high $\alpha_S=1.2$.
It is these regimes of negative (or to a lesser extent zero) 
charge which could in principle
lead to disastrous consequences if a negative strangelet were formed in
ultrarelativistic heavy ion collisions and began to grow.

However, the inclusion of finite size effects rules out such a disaster
scenario, because they reduce the number of negatively charged
$s$-quarks in the surface layer sufficiently to render the total $Z/A$
positive, as illustrated in Figures 2--4.

Quantitatively, stable strangelets are electrically positive for
radii from a few fm up to $10^6$ fm ($10^2<A<10^{18}$) for
$\alpha_S=0.9$, and for $10^2<A<10^7$ for $\alpha_S=1.2$. This is seen from
the detailed numerical calculations, but can be understood qualitatively
from the expression for $n_{s,S}$, which is the main contributor to the
change in charge for $A>10^2$. Thus $\Delta Z\approx -(1/3) n_{s,S} S \approx
0.18(\mu_s R)^2$ (the number assuming maximum depletion of $s$-quarks,
but the actual depletion is close to this for a wide range of
$\lambda_s$). With $A\approx V\mu_d^3/\pi^2$ and inserting typical
numbers for chemical potentials, this leads to 
\begin{equation}
\Delta Z/A \approx 0.3 A^{-1/3} \approx 0.3 R_{\rm fm}^{-1},
\end{equation}
in close agreement with the results of the full numerical calculations.

The numerical results depend on the choice of renormalization, but
qualitatively similar results are found for other choices of $\sigma$,
including the original\cite{freebal} 
(but physically questionable\cite{farhi84}) choice of $\sigma=m_s$.
A perturbative expansion to first order in $\alpha_S$ is unreliable at
high $\alpha_S$. In particular, the scheme used here breaks down at
$\alpha_S=\pi /2$, where light quark pressures become negative, but in
practice results for even somewhat lower $\alpha_S$ should be treated
with caution. One might also worry that the finite size terms
being known only to zeroth order could be inconsistent, but since these
terms are perturbations on the bulk solution itself, this omission
should matter only at order $\alpha_S^2$. 

But in spite of these reservations, the general qualitative result
demonstrated here in quantitative detail to first order in $\alpha_S$
within the MIT bag model, namely that surface depletion of $s$-quarks
removes sufficient negative charge to make strangelets positive below
some high $A$-value, seems quite robust. It
depends only on the fact that $f_{s}$ is negative, which leads to a
negative $n_{s,S}$, and a positive surface tension. This would seem a
natural condition in any model for stable strangelets\cite{nuclear}.

Thus the conclusion is that even within the limited range of parameters
where bulk strange quark matter is stable and negatively charged,
intermediate mass strangelets have positive quark charge because of
surface depletion of $s$-quarks. Even if very small strangelets might
still be created with negative charge in ultrarelativistic heavy ion
collisions due perhaps to stabilizing shell effects \cite{shell}
(such effects would
play no practical role at high $A$), they would not be able to
bridge the positively charged regime of intermediate $A$ on their way to
the putative state of bulk, negatively charged strange quark matter.
This rules out the disaster scenarios discussed recently.

I thank Bob Jaffe and Jack Sandweiss for comments. This work was
supported in part by the Theoretical Astrophysics Center under the
Danish National Research Foundation.

\begin{figure}
\epsfxsize=8.5cm\epsfbox{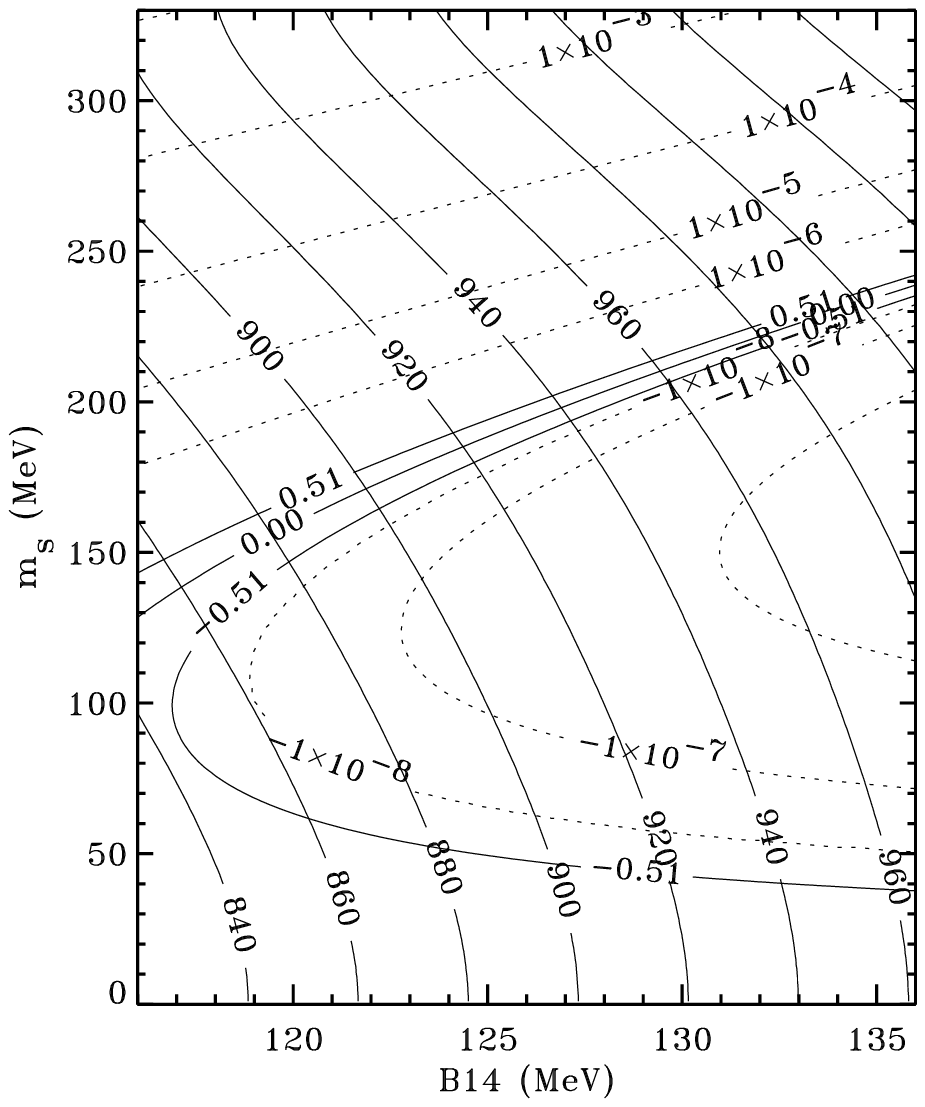}
\caption{Solid curves going from upper left to lower right show
constant energy per baryon of bulk strange quark matter in MeV as a
function of $B^{1/4}$ and $m_s$ for $\alpha_S=0.9$. 
$B$ is bounded from below by requiring
that $E/A>930$ MeV for up-down quark matter ($m_s \rightarrow\infty$).
Dotted curves give total number of electrons per baryon
(equal to the net quark charge per baryon).
Solid curves labeled 0.51, 0, and $-$0.51 indicate limits
where the electron chemical potential equals $m_e$, 0, and $-m_e$.}
\label{fig1}
\end{figure}

\begin{figure}
\epsfxsize=8.5cm\epsfbox{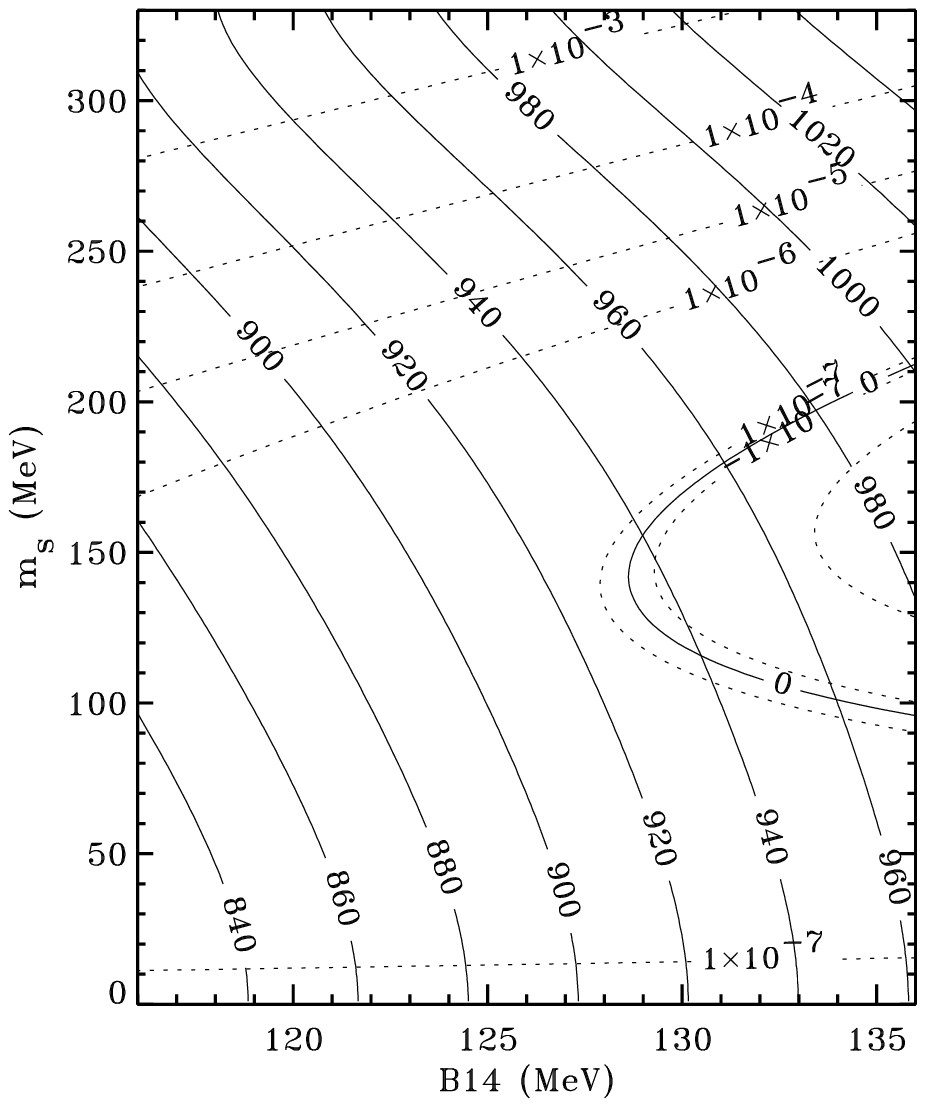}
\caption{As Figure 1, but for a strangelet with radius $10^6$fm,
corresponding to $A\approx 10^{18}$. Negative quark charge is now
limited to the small region in the middle right part of the diagram.
For systems smaller than this,
finite size effects make the total quark charge positive for all stable
strangelets ($E/A<930$ MeV).}
\label{fig2}
\end{figure}

\begin{figure}
\epsfxsize=8.5cm\epsfbox{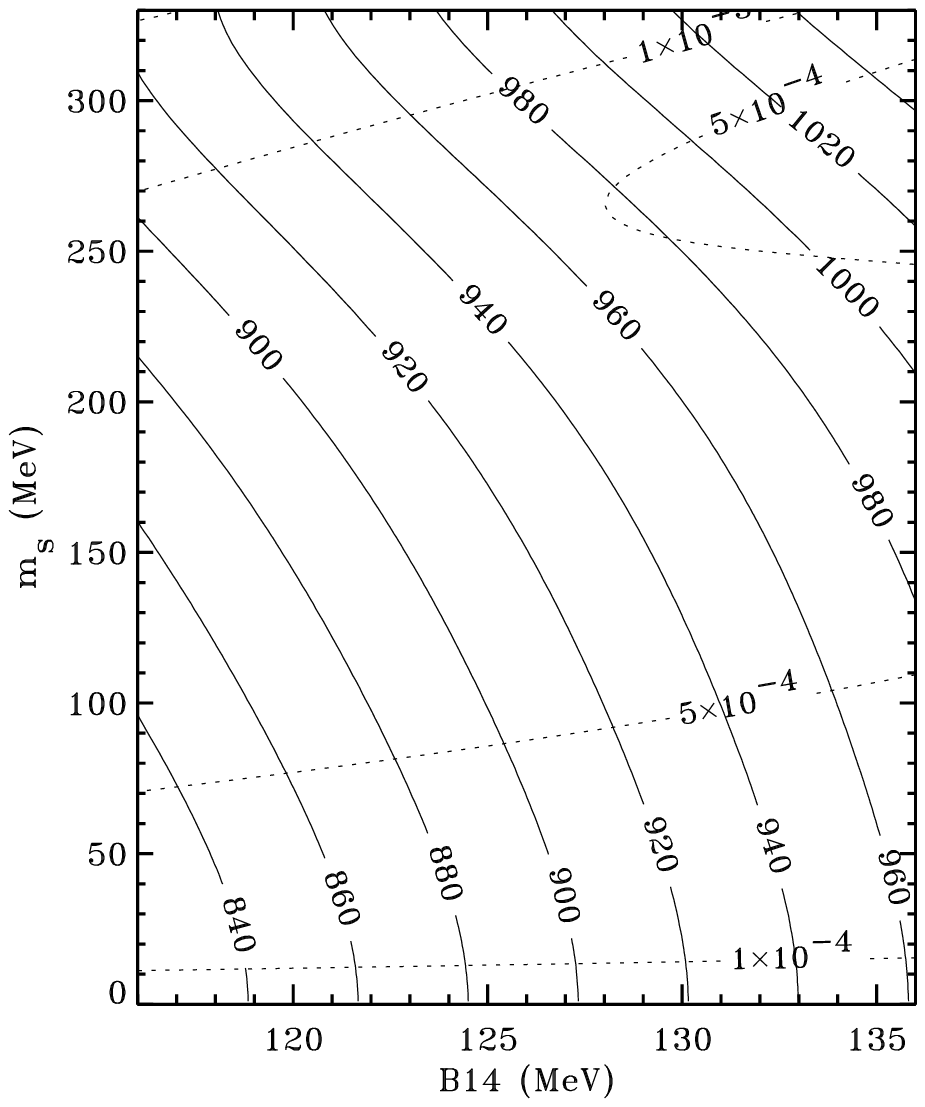}
\caption{As Figure 2, but for $R=1000$ fm ($A\approx 10^9$).
The total quark charge is positive throughout the diagram, typically
with $Z/A\approx 5\times 10^{-4}$.}
\label{fig3}
\end{figure}

\begin{figure}
\epsfxsize=8.5cm\epsfbox{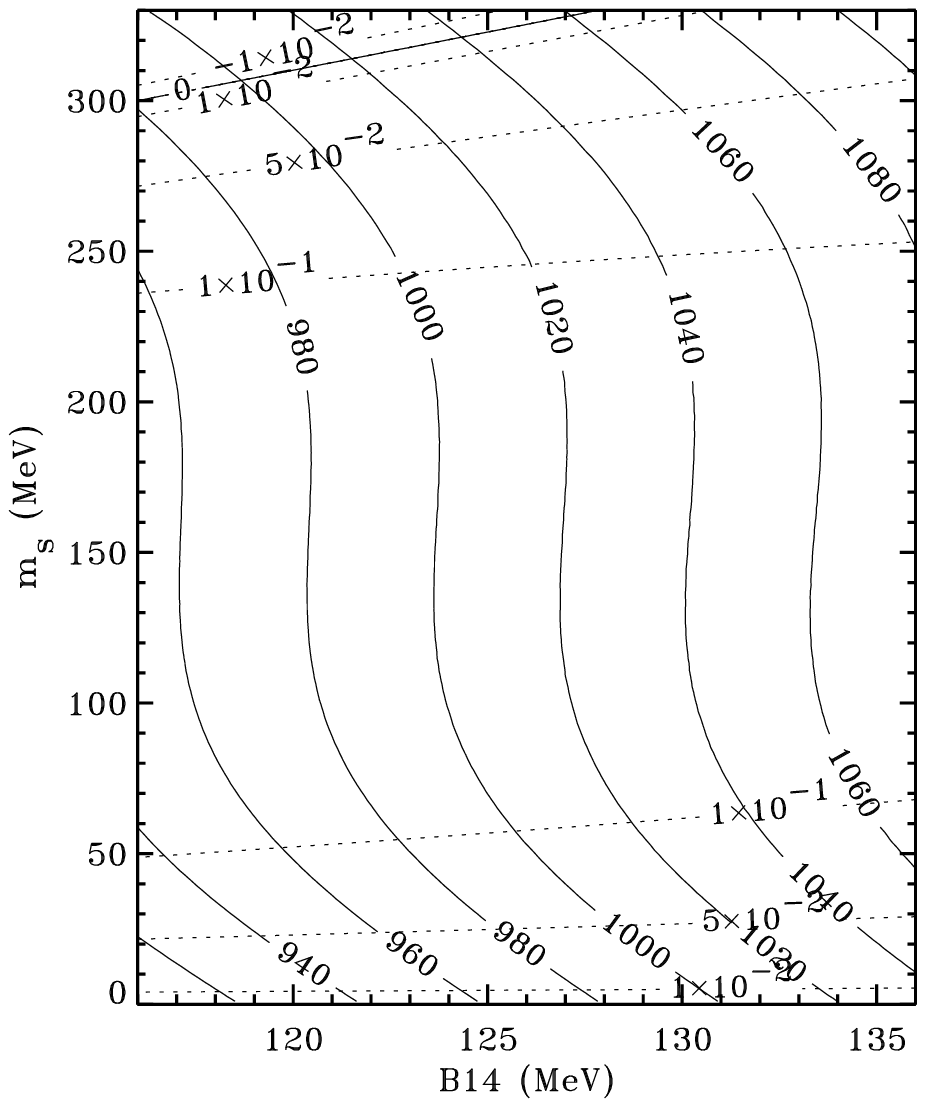}
\caption{As Figure 2, but for $R=3$ fm ($A\approx 30$).
$Z/A$ is positive almost everywhere, but a region of negative strangelet
charge appears for high $m_s$, caused mainly by the increasingly
important curvature contributions. Notice the significant increase in
energy per baryon typical of small strangelets because of surface and
curvature energy contributions. These small strangelets are 
at most metastable.}
\label{fig4}
\end{figure}


\begin{references}
\bibitem{charge} The matter is kept locally charge neutral by electrons
(positrons).
Finite size strangelets are surrounded by a thin atmosphere of
electrons (positrons), below which the quark ``core'' has a net positive
(negative) charge.
\end{references}
\end{document}